\pdfoutput=1
\RequirePackage{fix-cm}
\documentclass[twocolumn]{svjour3}          % twocolumn
\smartqed  % flush right qed marks, e.g. at end of proof
\usepackage{graphicx}
\usepackage{url}
\usepackage{xcolor,soul}
\usepackage[inline]{enumitem}
\def\makeheadbox{\relax}

\usepackage{array}
\newcolumntype{P}[1]{>{\centering\arraybackslash}p{#1}}
\newtheorem{Example}{Example}
\newtheorem{defn}{Definition}

\begin{document}

\title{Dataset search: a survey}


\author{Adriane Chapman        \and
        Elena Simperl \and
        Laura Koesten \and
        George Konstantinidis \and
        Luis-Daniel Ibáñez-Gonzalez \and
        Emilia Kacprzak \and
        Paul Groth
}


\institute{A. Chapman \at
              University of Southampton \\
              \email{adriane.chapman@soton.ac.uk}           %  \\
           \and
        E. Simperl \at
        University of Southampton \\
        \email{E.Simperl@soton.ac.uk}
        \and
        L. Koesten \at
        The Open Data Institute\\
        \email{laura.koesten@theodi.org}
        \and
        G. Konstantinidis \at
        University of Southampton\\
        \email{G.Konstantinidis@soton.ac.uk}
        \and
        L. Ibáñez-Gonzalez \at
        University of Southampton\\
        \email{L.D.Ibanez-Gonzalez@soton.ac.uk}
        \and
        Emilia Kacprzak \at
        The Open Data Institute\\
        \email{emilia.kacprzak@theodi.org}
        \and
        Paul Groth \at
       University of Amsterdam\\
        \email{p.groth@uva.nl}
}

\date{Received: date / Accepted: date}

\maketitle

\begin{abstract}
Generating value from data requires the ability to find, access and make sense of datasets. There are many efforts underway to encourage data sharing and reuse, from scientific publishers asking authors to submit data alongside manuscripts to data marketplaces, open data portals and data communities. Google recently beta released a search service for datasets, which allows users to discover data stored in various online repositories via keyword queries. These developments foreshadow an emerging research field around dataset search or retrieval that broadly encompasses frameworks, methods and tools that help match a user data need against a collection of datasets. Here, we survey the state of the art of research and commercial systems in dataset retrieval.  We identify what makes dataset search a research field in its own right, with unique challenges and methods and highlight open problems.  We look at approaches and implementations from related areas dataset search is drawing upon, including information retrieval, databases, entity-centric and tabular search in order to identify possible paths to resolve these open problems as well as immediate next steps  that will take the field forward.
\end{abstract}

\section{Introduction}
Data is increasingly used in decision making: to design public policies, identify customer needs, or run scientific experiments \cite{GOHAR2018114,townsend:2013}.
For instance, the integration of data from deployed sensor systems such as mobile phone networks, camera networks in intelligent transportation systems (ITS) \cite{Kitchin:2014} and smart meters \cite{alahakoon:2016}) is powering a number of innovative solutions such as the city of London's oversight dashboard \cite{batty:2016}. Datasets are increasingly being exposed for trade within data markets \cite{Balazinska2013,Grubenmann:2018:FWD:3178876.3186002} or shared via open data portals \cite{ckan,hendler_2012,kassen_2013,linkedopendatacloud,opendatamonitor,ukopendata} and scientific repositories \cite{elsevier,dataverse}. Communities such as Wikidata or the Linked Open Data Cloud \cite{linkedopendatacloud} come together to create and maintain vast, general-purpose data resources, which can be used by developers in applications as diverse as intelligent assistants, recommender systems and search engine optimization. The common intent is to broaden the use and impact of the millions of datasets that are being made available and shared across organizations \cite{Borgman2012,Pasquetto2017,Wilkinson2016}. This trend is reinforced by advances in machine learning and artificial intelligence, which rely on data to train, validate and enhance their algorithms \cite{DBLP:journals/corr/abs-1811-03402}. In order to support these uses, we must be able to search for datasets. Searching for data in principled ways has been researched for decades \cite{codd1972relational}. However, many properties of \emph{datasets} are unique, with interesting requirements and constraints. There are many open problems across dataset search, which the database community can assist with. 

Currently, there is a disconnect between what datasets are available, what dataset a user needs, and what datasets a user can actually find, trust and is able to use \cite{Borgman2012,DBLP:journals/corr/abs-1811-03402,Stonebraker2018}. Dataset search is largely keyword based over published metadata, whether it is performed over crawls across the web \cite{googledevelopers,sansone:2017} or within organizational holdings \cite{hendler_2012,kassen_2013,thelwall_figshare}. There are several problems with this approach. Available metadata may not encompass the actual information a user needs to assess whether the dataset is fit for a given task \cite{DBLP:conf/chi/KoestenKTS17}. Search results are returned to the user based on filters that were appropriate for web-based information, but do not always transfer well to datasets \cite{GregoryGCSW17}. These limitations impact the use of the retrieved data - machine learning can be unduly affected by the processing that was performed over a dataset prior to its release \cite{Szegedy2013IntriguingPO}, while knowing the original purpose for collecting the data aids interpretation and analysis \cite{woodall:2015}. In other words, in a dataset search context, approaches need to consider additional aspects such as data provenance \cite{Buneman:2006:PMC:1142473.1142534,Green:2007:PS:1265530.1265535,DBLP:journals/vldb/HerschelDL17,lee:2017,DBLP:series/synthesis/2013Moreau,wylot2017storing}, annotations \cite{Ibrahim:2015:PAM:2723372.2749435,limaye_annotating_2010,Xiao:2015:EMG:2723372.2735355}, quality \cite{Rekatsinas:2014:CSF:2588555.2610504,Umbrich2015,zaveri_quality_2016}, granularity of content  \cite{DBLP:journals/corr/abs-1810-12423}, and schema \cite{alexe2011,Begoli:2018:ACF:3183713.3190662} to effectively evaluate a dataset's fitness for a particular use. The user does not have the ability to introspect over large amounts of data, and their attention must be prioritized \cite{bailis2017}. In other cases, a user's need may require integrating data from different sources to form a new dataset \cite{gentile:2016,Rekatsinas:2014:CSF:2588555.2610504}. Furthermore, using a dataset is constrained by licenses and terms and conditions, which may prohibit such integration, especially when personal data is involved \cite{Mork2010}.

In order to realize the full potential of the datasets we are generating, maintaining and releasing, there is more research that must be done. Dataset search has not emerged in isolation, but has built on foundational work from other related areas. In Section \ref{sec:background}, we outline the basic dataset search problem, and provide a quick review of the sub-areas that have influenced dataset search. Current commercial dataset search offerings are outlined in Section \ref{sec:implementation} while Section \ref{sec:researchsurvey} provides a survey of dataset search research. Finally, Section \ref{sec:openproblems} provides a synopsis of open problems in dataset search as well as related research that could be applied. Section \ref{sec:route} highlights a possible route to take steps to advance the field.

\section{Background} \label{sec:background}
\begin{figure*}
\begin{center}
  \includegraphics[width=150mm,scale=0.4]{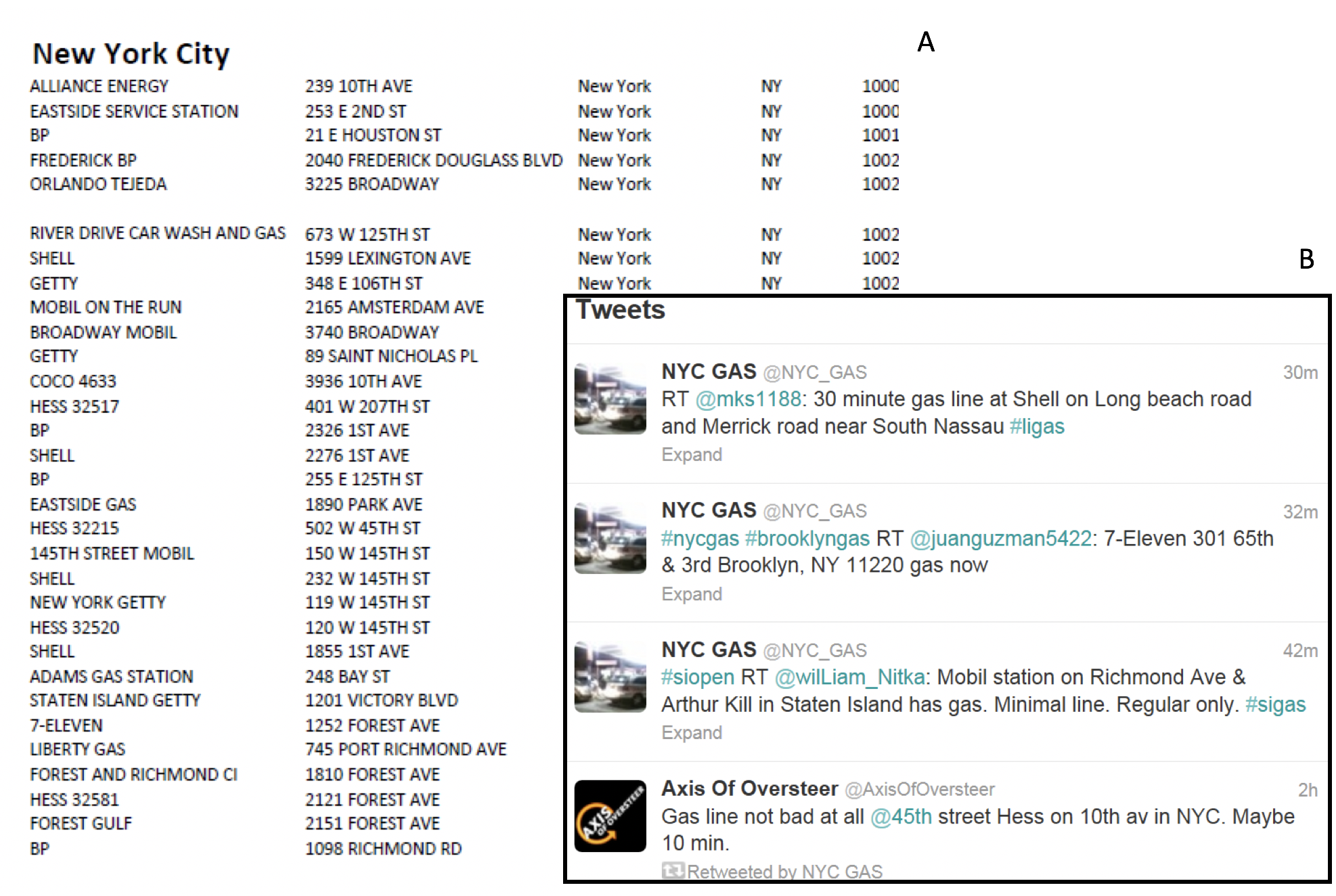}
\caption{Datasets about gasoline availability in New York City in the week after Hurricane Sandy in 2012. (a) The American Automobile Association (AAA) created a structured dataset twice post-Sandy by phoning every gas station in the NYC area. It is complete, easy to use (CSV), accurate, clean, and was out of date by the time it was released. (b) The second dataset is a collection of tweets to NYC\_GAS. It is incomplete, requires Natural Language Processing (NLP) techniques to use, is dirty with respect to place names and addresses, but is up to date and timely throughout post-hurricane clean-up efforts.}
\label{fig:exampledatasets}       % Give a unique label
\end{center}
\end{figure*}

To understand the fundamental problem of dataset search, we define a dataset. The concept of dataset is abstract, admitting several definitions depending on the particular community \cite{Borgman2012,Pasquetto2017}. There is a large body of work discussing the nature of data and its relation to practice and reuse \cite{Borgman2012,Borgman2015}. From a statistical point of view, the statistical data and metadata exchange initiative (SDMX) \cite{sdmxglossary} defines a dataset as \emph{`a collection of related observations, organized according to a predefined structure'}. This definition is shared by the DataCube vocabulary, which adds the notion of a \emph{`common dimensional structure'} \cite{datacube}. Meanwhile, the  Organization for Economic Co-operation and Development (OECD), citing the US bureau and census, uses \emph{`any permanently stored collection of information usually containing either case level data, aggregation of case level data, or statistical manipulations of either the case level or aggregated survey data, for multiple survey instances'} \cite{sdmxglossary}. The Data Catalog Vocabulary \cite{dcat} includes a dataset class, defined as a \emph{`collection of data, published or curated by a single agent, and available for access or download in one or more formats.'} Finally, for the  MELODA (MEtric for reLeasing Open DAta) initiative, a dataset is a \emph{`group of structured data retrievable in a link or single instruction as a whole to a single entity, with updating frequency larger than a once a minute'} \cite{meloda}. For the purposes of this paper, we will use the following definition:

\begin{defn} [Dataset]
A collection of related observations organized and formatted for a particular purpose.
\end{defn}

Thus, dataset search involves the discovery, exploration, and return of datasets to an end user. We note two very distinct types of dataset search in this work. In what we will call ``basic" dataset search, the set of related observations were organized for a particular purpose, and then released for consumption and reuse. We see this pattern of interaction within individual data repositories, such as Figshare \cite{thelwall_figshare}, Dataverse \cite{dataverse}, Elsevier Data Search \cite{elsevier}, Open Data Portals \cite{ckan,hendler_2012,kassen_2013,linkedopendatacloud,opendatamonitor,ukopendata} and global searches such as DataMed \cite{sansone:2017} or Google Dataset Search \cite{googleblog}. A basic search, using any of these services is discussed in Example \ref{ex:basic}. Alternatively, a dataset search may involve a set of related observations that are organized for a particular purpose by the searcher themselves. This pattern of behaviour is particularly marked in Data Lakes \cite{Gao:2018:NDL:3183713.3183746,dhsdatalake}, data markets \cite{Balazinska2013,Grubenmann:2018:FWD:3178876.3186002}, and tabular search \cite{Lehmberg:2017:SWT:3137628.3137657,Zhang2018AdHT}; Example \ref{ex:constructive} illustrates this kind of data search.

\begin{Example} [Basic dataset search] \label{ex:basic}
Imagine you want to write an article on how Hurricane Sandy impacted the gasoline prices in New York City in the week after the incident. Consider the two datasets shown in Figure \ref{fig:exampledatasets}. Dataset $A$ is from the American Automobile Association (AAA) and dataset $B$ is from Twitter, documenting the gasoline available for purchase in New York City in the week after Hurricane Sandy. The choice of which dataset to use depends on the specifics of the information need, potentially the purpose and requirements of algorithms or processing methods, as well as the user's tool-set and data literacy. In order to find the right dataset, a user must issue a query that will return datasets, not tuples, documents or corpora. Differences inherent in the datasets should alter their ranking. For instance, a user who requires easy-to-use data, with fewer restrictions on timeliness may feel that the AAA dataset is a better fit than the other one. A user who wishes to establish an accurate timeline of gas in NYC would have a different assessment. These two users have different purposes, and therefore would assess the datasets differently. Moreover, both users use the content (gasoline) as the initial inclusion requirement, but use very different criteria and metrics to rank the datasets. 
\end{Example}

%\begin{Example} [Cohesive Dataset Search]
%Consider the two datasets from the American Automobile Association (AAA) and Twitter that document the gasoline available for purchase in New York City in the week after Hurricane Sandy, shown in \ref{fig:exampledatasets}. When making a decision that requires information on gasoline in NYC, the choice of which dataset to use depends on the algorithms’ purpose and requirements, in addition to the users’ tool-set and technical strengths. In order to find the right dataset, a user must issue a query that will return datasets, not tuples, documents or corpora. Differences inherent in the datasets should alter their ranking. A user who requires easy-to-use data, with fewer restrictions on timeliness may feel that the quality of the AAA dataset is greater than the Twitter dataset. A user who wishes to establish an accurate timeline of gas in NYC would have a different quality assessment. These two users have different purposes, and therefore would rank the datasets differently. Moreover, both users use the content (gas) as the initial inclusion requirement, but use very different features and metrics to rank the datasets.
%\end{Example}

\begin{Example}[Constructive dataset search] \label{ex:constructive}
% Consider a local government that wishes to increase the number of households who are using solar energy instead of the power grid. In order to appropriately target offers of tax breaks, it must build a dataset of homes that have the appropriate directionality, dimensions, and identify whether they are primary residences. In order to do this, the following datasets are combined: aerial images to calculate roof directionality and dimension; city survey maps to identify addresses; tax records to identify residence status. 
In order to better understand the needs of the city, for instance to deal with flooding, the Centro De Operacoes Prefeitura Do Rio in Rio de Janeiro, Brazil mashes-up `traffic and public transport, municipal and utility services, emergency services, weather feeds, and information sent in by employees and the public via phone, internet and radio' \cite{Kitchin:2014}. Consider a simple scenario in which datasets on weather highlighting rain amounts that could trigger a flash flood are integrated on the fly with datasets on traffic volume and augmented with identification of emergency response services in order to create a dataset that highlights the current populations at risk during an event. A recent extension to RapidMiner highlights the opportunities inherent in creating a dataset, with additional examples \cite{gentile:2016}. 
\end{Example}

\subsection{Overview of generic dataset search}
\begin{figure}[t]
\begin{center}
  \includegraphics[width=85mm,scale=0.4]{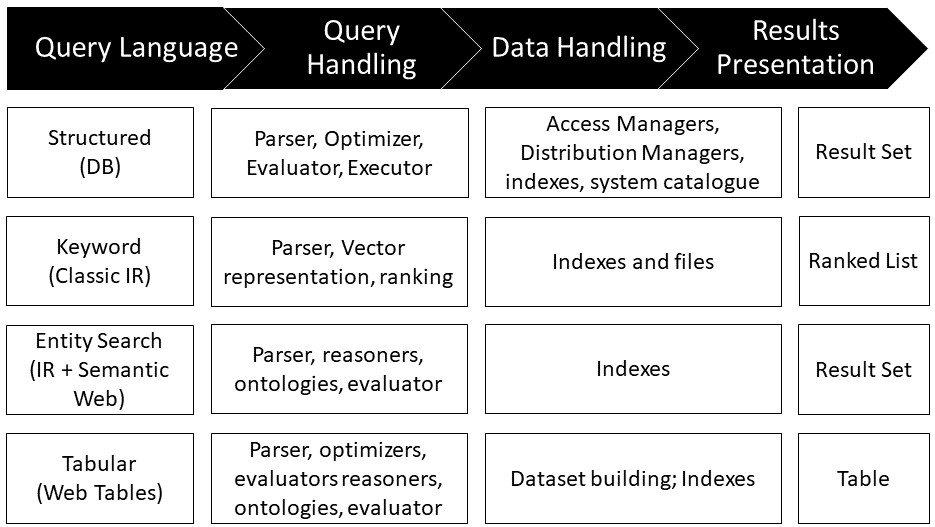}
\caption{An abstract view of the search process, comprising of querying, query processing, data handling and results presentation. Examples of how the database, IR, semantic web and tabular search communities implement these steps is shown.}    \label{fig:overview}   % Give a unique label
\end{center}
\end{figure}

 Figure \ref{fig:overview} contains a high-level view of the search process, as well as a mapping to other communities who are active in search. We will use the generalized steps indicated to outline the generic dataset search process below. A general approach to providing search over datasets is to model the user interface over existing keyword based information retrieval search systems where a user poses a query and a ranked list of existing datasets is returned. Indeed, a majority of data repositories provide this form of interface. 

\begin{figure}[t]
  \includegraphics[width=80mm,scale=0.4]{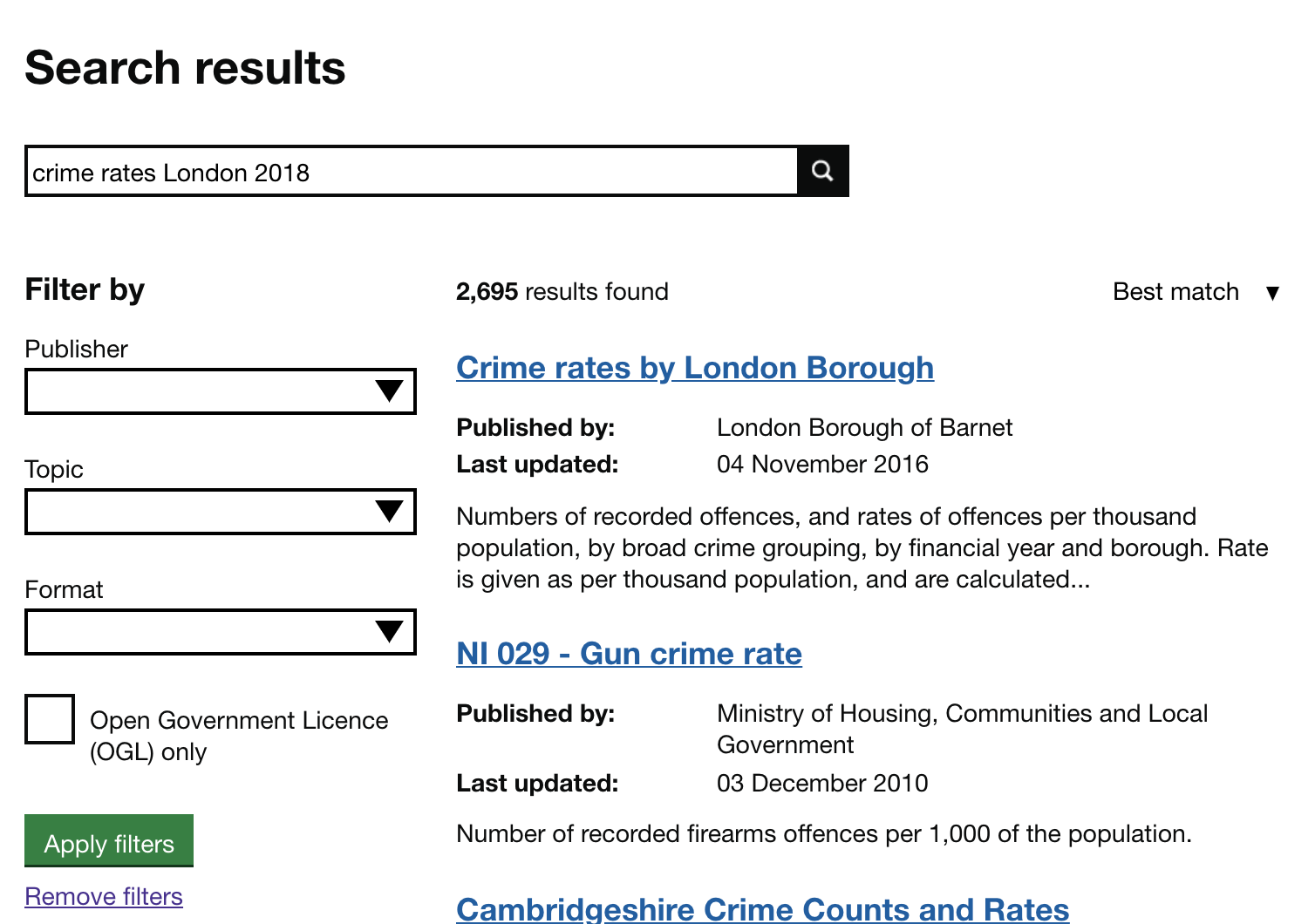}
\caption{Dataset search engine result page for the UK government's Open Data portal, data.gov.uk.}
\label{fig:datagovuk}       % Give a unique label
\end{figure}

\noindent \textbf{Querying.}  In the case of dataset search, a query is typically a keyword or Contextual Query Language (CQL) expression. Figure \ref{fig:datagovuk} shows the search interface for the UK government's Open Data portal \cite{ukopendata}. In addition to the keywords search box, the ``Filter by'' boxes allow the user to subset the data according to categories pre-identified by the repository. 

\noindent \textbf{Query Handling.} The keywords, and any categories, submitted by the user are used to search over the metadata published about a dataset. Based on the metadata similarity to the search terms, a result set is produced.

\noindent \textbf{Data Handling.} In preparation for querying, the dataset owners must populate the metadata about their dataset. For instance, the dataset publisher supplies information such as title, description, language, temporal coverage, etc.; DCAT \cite{dcat} is the W3C standard for interoperability of catalogues, and contains a representation and vocabulary for datasets. Additional metadata, such as summarizations \cite{DBLP:journals/corr/abs-1810-12423,nguyen2015result,Xiao:2015:EMG:2723372.2735355} could also be contributed. Unfortunately, the creation and maintenance of this metadata is currently resource intensive. 

\noindent \textbf{Results Presentation.} Search Engine's Results Pages (SERPs) for dataset search currently  follow a traditional \emph{10 blue links} paradigm, as can be seen on many data portals \cite{ukopendata,dataverse,hendler_2012,kassen_2013} as well as the Google Dataset search \cite{googledevelopers}. Basic filtering options, as can be seen in Figure \ref{fig:datagovuk}, are sometimes available for faceted search within specific portals. Clicking on a search result takes the user to a preview page that contains metadata, such as information about the publisher, publishing data, licensing, etc. (see DCAT \cite{dcat}). If available the preview page also contains a textual description. Many data portals will also include a preview by displaying a portion of the raw data  or a visualization of particular patterns.

\subsection{Common Search Architectures} As with searches in databases, IR and the semantic web, searches for datasets can be local, e.g. within a single repository \cite{elsevier,dataverse,thelwall_figshare,dhsdatalake}. In a similar manner to a distributed database, given a query $Q$ and a set of datasets (the \emph{sources}), the query engine first selects the datasets relevant to the query~\cite{saleem_hibiscus:_2014,vidal_selection_2016} and then chooses between different approaches: aggregating the datasets locally, using distributed processing as in Hadoop~\cite{wylot_rdf_2018}, or a federated approach~\cite{oguz_federated_2015}. 

The dataset search problem can be addressed at various levels. Services such as Google Dataset Search \cite{googledevelopers} and DataMed \cite{sansone:2017} crawl across the web and facilitate a global search across all distributed resources. These approaches use tags found in schema.org \cite{Guha:2016:SES:2886013.2844544} or DCAT \cite{dcat} to structure and identify the metadata considered important for datasets. However, the problem also exists at a local level, including open government portals such as data.gov.uk \cite{ukopendata},  organizational data lakes \cite{dhsdatalake}, scientific repositories such as Elsevier's \cite{elsevier} and data markets \cite{Balazinska2013,Grubenmann:2018:FWD:3178876.3186002}. Across all these systems, users are attempting to discover and assess datasets for a particular purpose. Supporting them requires frameworks, methods and tools that specifically targets data as its input form and the specific information needs of data professionals. 

\subsection{Search sub-communities} \label{sec:otherareas}
Search has been addressed across many computer science sub-disciplines, such as databases, information retrieval, entity-centric search and tabular search. Figure \ref{fig:overview} contains an overview of the high-level steps, and how each sub-discipline implements them. While dataset search is a subject in its own right, with distinct challenges and characteristics, it shares commonalities and draws upon insights from all these areas. In this section, we provide a very brief review of the focus and tools each community uses. We focus specifically on sub-areas in which the type of object returned is the same as the underlying data, e.g. a result set of data from a database of data, or a document from a corpus of documents. We neglect approaches such as question-answering \cite{Kwok:2001:SQA:502115.502117} in which extra reasoning and manipulation of the returned result set is possible.

\subsubsection{Entity-centric search}
The task of entity search is to provide information about a specific named-entity (e.g. person, place, organization, ...) \cite{Balog2010}. For a comprehensive view of entity-oriented search we refer the reader to \cite{balog2018entity}. Here, we introduce  the work from the semantic web community in this space as it focuses on entities represented in data and not derived from text. 

The semantic web community has worked towards creating machine-understandable graph-based representations of data \cite{ldbook}. It proposes languages, models and techniques to publish data online in the form of entities, properties, literals, and, most importantly, links to other resources. These links facilitate search and exploration of a global decentralized data space, similar to browsing and navigation on the web. The World Wide Web Consortium (W3C) settled on the Resource Description Format (RDF) as a standard model for representing and exchanging data about resources, which can refer to conventional web content as well as entities in the offline world such as people, places and organizations, identified by \emph{International Resource Identifiers} (IRIs). Properties link entities or attach attributes to them. By reusing and linking IRIs, publishers signal that they hold data about the same entity, therefore enabling queries such as \emph{Who holds data about England?} and \emph{What do you know about England?} across multiple resources without any additional integration effort. 

Realising the web of data requires several steps: lifting existing data into the semantic web, commonly referred to as linked (open) data \cite{ldcycle}, defining vocabularies and schemas to describe data in RDF and connecting to other datasets. For example, the Linked Open Vocabulary portal\footnote{\url{https://lov.linkeddata.es}} lists $635$ such vocabularies, and provides search and exploration capabilities to find commonly used IRIs, to assist publishers in reusing them to facilitate data interpretation and interlinking. 

Interlinking comprises two complementary problems. First is \emph{entity resolution}: given two or more datasets, identify which entities and properties are the same. A general framework of entity resolution is described \cite{erbook}. It covers the design of \emph{similarity metrics} to compare entity descriptions, and the development of \emph{blocking} techniques to group roughly similar entities together, in order to not apply more expensive similarity metrics to entities that have a low chance of ultimately resolving to the same entity. Finally, more recent efforts have tried \emph{iterative} approaches, where discovered matches are used as input for computing similarities between further entities. The second part of interlinking is referred to as \emph{link discovery}, where given two datasets, one has to find properties that hold between their entities. Properties can be equivalence or equality, as in entity resolution, or domain specific such as 'part-of' \cite{ldbook}.

Information that is interlinked in this way allows for entity-centric searching, by identifying entities in the query and where they have similar matches in the data \cite{Zhang2017EntiTablesSA,Zhang:2018:OTG:3209978.3209988}.

\subsubsection{Information Retrieval}
IR systems can be broadly classified in web and document search engines, and engines for other types of entities (images, people, etc), called \emph{vertical search engines}.
%Depending of the underlying resources information retrieval algorithms are adapted to maximise their efficiency. 
The web and document engines use a number of statistical techniques to compute the relevance of a word (coming from an user query) to a document. 

Vertical search engines are specifically tailored to the characteristics of the resources. For example, an email search has unique sets of  resources for each users in addition to specific metadata such as sender and receiver addresses, topic or timestamp in order to judge the relevance \cite{Ai:2017:CES:3038912.3052615}. Due to the specificity and limited scope of resources, vertical search engines often offer greater precision, utilize more complex schemas to match specific searching scenarios, and tend to support more complex user tasks \cite{li2008time,DBLP:conf/webi/YuLL05,DBLP:conf/sigir/WeerkampBKMBR11}. 

\subsubsection{Databases}
The classic pipeline for search within a database begins with a structured query. Once a query is issued, the classic pipeline includes: parsing the query; creating an evaluation plan; optimizing the plan; executing the plan utilizing appropriate indexes and catalogues. Other sources can provide greater details on each of these steps.

% The classic pipeline for search within a database begins with a structured query. Once a query is issued, the classic pipeline includes: parsing the query \cite{Chen:2000:NSC:342009.335432,Li:2005:NIN:1066157.1066281}; creating an evaluation plan \cite{Levy1995}; optimizing the plan \cite{Chaudhuri:1998:OQO:275487.275492,Ioannidis:1996:QO:234313.234367}; executing the plan utilizing appropriate indexes and catalogues \cite{bertino:2012}.

\subsubsection{Hidden/deep web}
The \emph{hidden}, or \emph{deep}, web refers to datasets that lie ``behind'' web forms typically written in HTML \cite{he2007accessing,madhavan2008google}, and ranging from medical research data to financial information and shopping catalogues. To access data behind a form a user needs to insert input text and submit the form, in order to be directed to a web page presenting the appropriate dataset \cite{khare2010understanding}. It has been estimated that the data stored in hidden databases is an order of magnitude larger than the so-called \emph{surface} web data, i.e., the data directly accessed by web crawlers ~\cite{he2007accessing,madhavan2008google}.

There have been two main approaches to searching for data on the deep web. The first uses more traditional techniques to build vertical search engines, whereby semantic mappings are constructed between each website and a centralized mediator tailored to a particular domain. Structured queries are posed on the mediator and redirected as appropriate queries over the web forms using the mappings.  Kosmix \cite{rajaraman2009kosmix} (later transformed into WalmartLabs.com) was such a system presenting vertical engines for a large number of domains, ranging from health, and scientific data to car and flight sales. Other approaches to build such vertical search engines and automatically align different web forms,  learn the forms' possible inputs, and create centralized mediated forms~\cite{he2007accessing}.    
 A second group of approaches tries to generate the resulting web pages, usually in HTML, that come out of web form searches.  Google has proposed a method for such \emph{surfacing} of deep web content by automatically estimating input to several millions of HTML forms, written in many languages and spanning over hundreds of domains, and adding the resulting HTML pages into the Google search engine index \cite{madhavan2008google}. The form inputs are stored as part of the indexed URL, and when users click on a search result they are directed to the result of the (freshly submitted) form. 
 
 \subsubsection{Tabular Search}
 In many cases, users are not interested in finding one specific dataset but instead are interested in extending or filling out an existing dataset, usually in the form of a table. Imagine a table with a series of columns and a series of rows, broadly the aim is to add additional relevant rows, columns or to fill in missing cell values. Thus, the input to the process is a table and the corresponding output is an enriched table. \cite{Yakout:2012:IEA:2213836.2213848} identified three core tasks in the augmentation of tables.

\begin{enumerate}
    \item Augmentation by attribute name - given a populated table and a new column name (i.e. attribute), populate the column with values. This is also referred to \emph{table extension} elsewhere \cite{Cafarella:2008:WEP:1453856.1453916}. One can see this as finding tables which can be joined.
    \item Attribute discovery - given a populated table, discover new potential column names.
    \item Augmentation by example - given a populated table where some values are missing, fill in the missing values. This often referred to as \emph{table completion} in the literature~\cite{Zhang2017EntiTablesSA}. This task is like finding tables which can be unioned. 
\end{enumerate}

We refer to the combination of table extension, table completion, and attribute discovery as \emph{tabular search}. This highlights that the query itself is a table and lends itself to the information retrieval perspective where the challenge is to answer the latent information need of the user. It is important to distinguish this task from \emph{table search} which is the discovery of tables given a keyword search. Table search is a sub-task of dataset search. 

\noindent {\bf Table extension}
\cite{LEHMBERG2015159} is divided into constrained and unconstrained table extension. Constrained table extension is essentially the augmentation by attribute name previously defined. Unconstrained table extension is also the addition of additional columns to a table but with no predefined label for the attribute. One can think of this as attribute discovery followed by constrained table extension. 

A common technique to perform table extension is to discover existing tables through table similarity - in particular by measuring \emph{schema similarity} \cite{das2012finding}. Indeed, table extension was introduced by \cite{Cafarella:2008:WEP:1453856.1453916} where they defined a special operator EXTEND that would discover similar web tables to the given input table. Similarity here is computed with respect to the schema of the table. The values of the most similar table are then used to populate the input table's additional column.  The Infogather system~\cite{Yakout:2012:IEA:2213836.2213848} uses a similar approach but instead of just calculating the direct similarity between the input table and potential augmenting tables it also takes into the account the neighborhood around the potential augmenting tables. These indirect tables provide ancillary information that can be better suited for augmentation than the tables with the highest similarity to the input tables. Of interest, \cite{das2012finding} have discovered that with respect to web tables there seems to be a latent link structure between tables. Recent work in table similarity has shown that semantic similarity using embedding approaches can improve performance over syntactic similarity measures \cite{Zhang2018AdHT}.

\noindent {\bf Table Completion} also relies heavily on table similarity as the mechanism for finding potential values that can be added to a table. \cite{Zhang2017EntiTablesSA} defines the notion of row population, which adds additional rows to a table. For simplicity, we view this as a type of table completion in which the values to be complete form an additional row. Even more broadly, one could provide a set of columns as a query and have the system fill in the remaining rows \cite{Pimplikar:2012:ATQ:2336664.2336665}. 

The task of table completion can be seen as entity set completion where the goal is to complete a list given a set of seed entities \cite{das2012finding,Zhang2017EntiTablesSA}. This task is relevant for a number of other tasks beyond table completion, including entity search \cite{Balog2010OverviewOT} and knowledge base completion~\cite{Dalvi:2012:WES:2124295.2124327}.
The completion of rows is similar to the broad problem of imputation and dealing with incomplete data \cite{miao2018incomplete}. Specific work in the context of the web has looked at performing imputation through the use of external data \cite{ahmadov2015towards,li2014web,tang2017efficient}. Much of that work has used web tables as the data source.

\section{Current implementations} 
\label{sec:implementation}
There are many functioning versions of dataset search in production today. In this section, we break down the set of dataset search services that exist according to their focus and how they deal with datasets.

\subsection{The Google Model} 

% With the growth of the number of the datasets available on the web and in enterprise repositories, different solutions have been proposed in order to aid users to retrieve this data. For example  Cafarella et al. propose targeting structured data (HTML tables) within the web pages and utilising the general web ranking with additional features to describe the tables within the web page \cite{Cafarella:2008:WEP:1453856.1453916}. The result, Fusion Tables, is a data management system that contains an HTML page for each table. The HTML page can be crawled and indexed in the same way as every other web page. Below we outline the currently applied solutions to the problem of dataset retrieval and the challenges that they face with datasets.

\noindent {\bf Google's systems.} In $2016$, Google introduced  \textit{Goods}, an enterprise dataset search system, to manage datasets originating from different departments within the company with no unified structure or metadata \cite{halevy2016goods}. In this catalog, related datasets are clustered based on the structure of the dataset or gathering frequency. Members of a group then become a single entry in the catalog. This helps to structure the catalog and also reduces the workload of metadata generation and schema computing. Within the Goods system each dataset entry has an overview of the dataset presented on a profile page. Using this profile, users can judge the dataset's usefulness to their task. Keyword queries are then laid on top of this structure, producing a ranked result list of datasets as an output. Search functionality was built based on an inverted index of a subset of the dataset’s metadata. In the absence of the information on the importance of each resource, \cite{halevy2016goods}  propose to rank the datasets based on heuristics over the type of a resource, precision of keyword match, if the dataset is used by other datasets and if the dataset contains an owner-sourced description. %They also point out that those heuristics are incomplete.

Following this work, in $2018$ Google introduced a vertical web search engine tailored towards datasets on the web \cite{googleblog}.
This system uses schema.org \cite{Guha:2016:SES:2886013.2844544}, which is a schema for describing structured data on the web, and is applicable across a wide variety of data formats. It can be used as markup to describe structured content (e.g. tables within web pages) or as a metadata schema describing specific data with a defined list of metadata attributes. %- for example a dataset\footnote{\url{http://schema.org/Dataset}}.
Google crawls the web for all the datasets described with use of schema.org \textit{Dataset} class and collects all the metadata provided to describe a given resource.
They further build the search capabilities on top of this metadata with additional information such as PageRank score of a page which contains metadata describing a given dataset \cite{googledevelopers}. 

\noindent {\bf Open data portals.} Like Google, the open data portals \cite{ckan,linkedopendatacloud,opendatamonitor,ukopendata,hendler_2012,kassen_2013} provide search over the metadata of available datasets. The most popular platform in the governmental open data domain is CKAN \cite{ckan}.
CKAN is built using Apache Solr\footnote{\url{http://lucene.apache.org/solr/}}, which uses Lucene to index the documents. In this scenario, the documents are the datasets' metadata provided by the publishers. CKAN integrates the DCAT metadata schema which is an RDF vocabulary facilitating interoperability between data catalogs published on the web \cite{dcat}. The main difference is that the open data portals do not need to crawl to collect this metadata. The open data portals catalogue their resources into pre-specified categories such as filetype, geographic region, etc. In addition metadata descriptions according to standards such as DCAT, which defines attributes such as title, description, language or licence \cite{dcat} are also maintained.
Despite the search functions provided by such catalogues, it is often not possible for an ordinary user to find relevant pieces of information quickly.  This can be caused by: non-intuitive or limited data descriptions; misleading naming conventions; incorrect assignment of categories to datasets; the user’s lack of in-depth knowledge of the subject; or simply because the search is only conducted over the metadata records provided by the publishing bodies rather than the data itself \cite{GregoryGCSW17}. The metadata describing datasets is often incomplete or outdated, as maintaining it is frequently manual and expensive. In many cases the metadata does not describe the full potential of the data, so some relevant datasets may not be presented as a result of a query simply because appropriate keywords were not used in the description.

In addition to DCAT \cite{dcat} and Schema.org \cite{Guha:2016:SES:2886013.2844544}, other efforts were introduced to accommodate the most popular data format on the web. For example the `CSV on the Web' working group has developed a standard for expressing useful metadata about tabular resources and CSV files specifically \cite{Tennison:16:CWA}. 
Their goal is to provide a uniform way of ensuring consistency of data types and formats (e.g. uniqueness of values within a single column) for every file, which can provide basis for validation and prevent potential errors.

\subsection{The Adding Value Model}	
In order to be more useful to a specific set of end users, many domains have also adopted strategies to effectively curate the contents of their search results for their specific end users. The searches for domain-specific datasets have corrallaries to both the vertical web search engine provided by Google and the in-house searches of the Open Data portals. For instance, DataMed, a biomedical search engine uses a suite of tags, DATS, to allow a crawler to automatically index scientific datasets for search \cite{sansone:2017}. The Open Contracting Partnership released a Open Contracting Data Standard that identifies information needed about contracts to allow their crawler to access and catalogue contracting datasets \cite{opencontractingstandard}.  On the other hand, data repositories like Elsevier \cite{elsevier}, Figshare \cite{thelwall_figshare}, Dataverse \cite{dataverse}, and many Open Data portals \cite{ukopendata,hendler_2012,kassen_2013}, have no need of crawling, and primarily search over metadata contained within their purview. 

The common theme of current dataset search strategies, both on the web and within the boundaries of a repository, is the reliance on  dataset publishers tagging their data with appropriate information in the correct format. Because current dataset search only uses the metadata view of a dataset, it is imperative that these metadata descriptions are correct and maintained. Other, domain-specific solutions function in similar ways. 

In aid of better searches, there are several attempts at monitoring and working over Open Data portals to provide a meta-analysis. For instance, the Open Data Portal Watch \cite{neumaier2018enabling,Neumaier:2016:AQA:3012403.2964909} currently watches $261$ open data portals. Once a week, the metadata from all watched portals is fetched, the quality of the metadata computed, and the site updated to allow an cohesive search across the open data.  Similarly, the Open Data Monitor reviews open data portals, and identifies where to search for information, in addition to assisting data owners successfully open their data \cite{opendatamonitor}. 

\subsection{The Constructive Dataset Model} 
Many private companies have understood that data is a commodity that can be effectively monetized. Some companies, such as Thomson Reuter have been collecting data to create datasets for sale for decades\footnote{\url{https://www.thomsonreuters.com/}}. However, companies such as OpenCorporates uses public data sources, with provenance, to gather information on legal entities. This dataset is then made publicly available\footnote{\url{opencorporates.com/}}. Similarly, Researchably compiles information from scientific publications and makes interest-specific datasets for sale to biotech companies\footnote{\url{https://www.researchably.com/}}. In all of these cases, the data exists in a scattered manner, and the company provides value by gathering, organizing and releasing it as a constructed dataset. 

\noindent {\bf Data Markets} exist as a way for organizations to realize value for their data \cite{Balazinska2013,GregoryGCSW17,Grubenmann:2018:FWD:3178876.3186002}. While the user is able to download the entire dataset from a data market, it is also possible to access subsets of the data as needed to construct a dataset.
 
\section{Survey of Dataset Search Research} \label{sec:researchsurvey}
This section surveys the current work related to dataset search. To organize it, we utilize the headings from Figure \ref{fig:overview}.
\subsection{Querying} 

\noindent \textbf{Creating queries.} Users interact with datasets in a different manner than they interact with documents \cite{kern2015}. While this study is limited to social scientists, it indicates that users have a higher investment in the results, and are thus willing to spend more time searching. Moreover, the relationship of the dataset to the task at hand may play a larger role in dataset search; e.g. two datasets about cars could fit within a user's ability to understand and utilize, but may have very different results when Data-centric tasks can be categorized into two  categories: (1) \emph{Process-oriented tasks} used to produce an end analysis and
(2) \emph{Goal-oriented tasks} used in a machine learning process \cite{DBLP:conf/chi/KoestenKTS17}.  While the boundaries between the two categories are somewhat fluid and the same user might engage in both types of tasks, the primary difference between them lies in the `user information needs', i.e. the details users need to know about the data in order to interact with it effectively. For process-oriented tasks, aspects such as timeliness, licenses, updates, quality, methods of data collection and provenance have a high priority. For goal-oriented tasks, intrinsic qualities of data such as coverage and granularity play a larger role.  As yet, beyond the user filtering by certain characteristics, there is no way to state the task needs in the query. There has not yet been a movement away from keywords and CQL to query datasets.

\noindent \textbf{Query Types.} As stated earlier, most queries for datasets use keywords or CQL over the metadata of the dataset. A formal query language that supports dataset retrieval does not yet exist. Instead, specific query interfaces are created for the underlying data type, e.g.  \cite{jain2007sql}  provides a SQL interface over text data and \cite{neumaier2018enabling} for temporal and spatial data. Current implementations provide platform specific faceted search to allow basic filtering for categories such as publisher, format, license or topics (for instance \cite{ukopendata}).

\subsection{Query Handling} 
 As stated in Section \ref{sec:background}, most dataset searches operate over the dataset's metadata. Unfortunately, low metadata quality (or missing metadata) affects both the discovery and the consumption of the datasets within Open Data Portals \cite{Umbrich2015}.  The success of the search functionality depends on the publishers knowledge of the dataset and the quality of the descriptions they provide. 

Moving away from just searching over the metadata, \cite{DBLP:conf/adcs/ThomasOR15} use the data type and column information for mapping columns in a query to the underlying table columns, while \cite{Pimplikar:2012:ATQ:2336664.2336665} allow keyword queries over columns. Similarly, \cite{gupta_karma:_2012} describe how to map structured sources into a semantic search capability. This is taken further in \cite{Zhang2018AdHT} by providing the ability to pose a keyword query over a \emph{table}.  

\subsection{Data Handling} 
While the ``handling" that typically needs to occur for dataset search at the moment is collection and indexing of metadata, there is research in additional data handling that can improve the effectiveness of search. 

\noindent\textbf{Quality and Entity Resolution.}
There are several efforts dealing with metadata quality \cite{Neumaier:2016:AQA:3012403.2964909,Umbrich2015}. One solution proposed to tackle the metadata quality problem include cross-validating metadata by merging feeds from identified entities \cite{heyvaert2015merging}. Using self-categorized information \cite{kunze_dataset_2013} as facets is another.  Attempts to better represent the underlying data \cite{BISCHOF201822} do have an affect on search. This includes better links with others data \cite{ellefi2016dataset}.

In the context of constructive dataset search, the Mannheim Search Join Engine \cite{Lehmberg:2017:SWT:3137628.3137657,LEHMBERG2015159} and WikiTables \cite{bhagavatula2013methods} use a table similarity approach for table extension but also look at the unconstrained task. In both cases, a similarity ranking between the input and augmentation tables is used to decide which columns should be added. Interestingly, the Mannheim system also consolidates columns from different potential augmentation tables before performing the table extension. 

\noindent\textbf{Summarization and Annotations.} To help both search and user understanding, summarizations and annotations are additional metadata that can be generated about the underlying dataset \cite{DBLP:journals/corr/abs-1810-12423}. For instance, \cite{Mork2010} deal with the problem that the underlying dataset cannot be exposed, but good summaries may help the user undertake the task of data access. Meanwhile, \cite{limaye_annotating_2010} use annotations to help support searching over data types and entities within a dataset, while \cite{kacprzak2018making} provide better labeling for numerical data in tables.

\subsection{Results Presentation} 
\noindent\textbf{Ranking Datasets.} There are several works that look at ranking datasets. Of the most basic, after performing a keyword query over tables, a ranking on the returned tables is attempted \cite{Zhang2018AdHT}. In a more advanced method, \cite{van-gysel-neural-2018} use an unsupervised learning approach to identify topics of database that can then be used in ranking. Finally, \cite{Li:2010:RCP:1920841.1920923} rank datasets containing continuous information.

% There are several actions a user may perform in order to better understand the data presented in a result set \cite{DBLP:conf/chi/KoestenKTS17}, including: (1) \emph{linking}; (2) \emph{time series analysis}; (3) \emph{summarizing}; (4) \emph{presenting}; and (5) \emph{exporting}.
% \emph{Linking} is about finding commonalities and differences between two or more datasets.  
% In a \emph{Time series analysis} data is ordered by time. The aim is to identify trends, or detect and predict events \cite{chatfield2016analysis}. \emph{Summarizing} involves creating a more compact, meaningful representation of the data. 
% \emph{Presenting} includes activities that transform data into human-friendly formats, such as visualize them or producing textual descriptions of the data. Finally, \emph{exporting} refers to all aspects around producing and publishing a dataset in a given format, including metadata. 

\noindent\textbf{Interactions.} Interactive query interfaces allow ad-hoc data analysis and exploration. Facilitating users exploration changes the fundamental requirements of the supporting infrastructure with respect to processing and workload \cite{Jiang:2018:EID:3183713.3197386}. 
Choosing a dataset greatly depends on the information provided alongside it. A number of studies indicate that standard metadata does not provide sufficient information for dataset reuse \cite{DBLP:journals/corr/abs-1810-12423,nguyen2015result}.  Recent studies have discussed textual (\cite{DBLP:journals/corr/abs-1810-12423,DBLP:conf/adcs/ThomasOR15}) or visual \cite{wiggins2018exploring} surrogates of datasets that aim to help people identify relevant documents and increase accuracy and/or satisfaction with their relevance judgments. 

There has been additional research in how to help users interact with datasets for better understanding. For instance, there is the \textit{many-answer problem}: users struggle to specify exact queries without knowing the data and their need to understand what is available in the whole result set to formulate and refine queries \cite{DBLP:conf/sigmod/LiuJ09}. Currently dataset search is mainly performed over metadata, so the users understanding of what the dataset contains before download is limited by the quality, comprehensiveness and nature of metadata. A number of frameworks or SERP designs have been proposed as research prototypes for data search and exploration, such as TableLens (\cite{Pirolli:1996:TLT:948449.948460}, DataLens \cite{DBLP:conf/sigmod/LiuJ09}, the relation browser \cite{marchionini2005} for sensemaking with statistical data, or summarization approaches of aggregate query answers in databases \cite{Wen:2018:ISE:3275366.3284965}. Navigational structures can support the cognitive representation of information \cite{rieh2016towards} and we see a large space to explore interfaces that allow more complex interaction with datasets such as sophisticated querying \cite{DBLP:conf/sigmod/JagadishCEJLNY07} (e.g. taking a dataset as input and searching for similar ones) or being able to follow links between entities in datasets. 

Interaction characteristics for dataset search have been subject to several recent human data interaction studies. Moving beyond search as a technological problem, \cite{GregoryGCSW17} show that there are also social considerations that impact a user when searching. In a comparison between document retrieval and dataset retrieval, \cite{kern2015} show that users are more reliant on metadata when performing dataset search. While looking at dataset users of varying abilities \cite{Boukhelifa:2017:DWC:3025453.3025738} show that the amount to tool support can impact a user's ability to effectively discover and use a dataset. Finally, in a framework for Human Interaction with Structured data 
\cite{DBLP:conf/chi/KoestenKTS17} discuss three major aspects that matter to data practitioners when selecting a dataset to work with: \textit{relevance}, \textit{usability} and \textit{quality}.
Users judge the relevance of datasets for a specific task based on the dataset's scope (e.g. geographical and temporal scope)
%\cite{neumaier2018enabling,DBLP:conf/www/KacprzakKTS18}),
\cite{neumaier2018enabling,kacprzak2018},
basic statistics about the dataset such as counts and value ranges, and information about granularity of information in the data \cite{DBLP:journals/corr/abs-1810-12423}. The documentation of variables and the context from which the dataset comes from also play a key role. Data quality is intertwined with a user's assessment of  ``fitness for
use"  and  depends on various factors (dimensions or characteristics) such as accuracy, timeliness, completeness, relevancy, objectivity, believability, understandability, consistency, conciseness, availability and verifiability \cite{DBLP:journals/corr/abs-1810-12423}. Provenance is a prevalent attribute to judge a datasets quality as it gives an indication of the authoritativeness, trustworthiness, context and original purpose of a dataset, e.g. \cite{DBLP:journals/corr/abs-1810-12423,DBLP:series/synthesis/2013Moreau}.  In order to judge a dataset's usability for a given task, the following attributes have been identified as important: format, size, documentation, language (e.g. used in headers or for string values), comparability (e.g., identifiers, units of measurement), references to connected sources, and access (e.g. license, API) \cite{DBLP:journals/corr/abs-1810-12423}. These are attributes independent of a dataset's content or topical relevance which can influence whether a user is actually able to enagage with a dataset.

% \subsection{User and task}\label{sec:userandtask}
%  Dataset fitness is task dependent, in that even if a user can find and understand a dataset, it may still be unsuitable for a specific task. This is, to an extent, subjective, as reflected in the digital literacy and trust quadrants of Figure \ref{fig:1}, and related to intended-use or context. This user-based subjectivity requires interdisciplinary methods to fully investigate. For instance, recent work attempts to peel apart how users search out data, and seek to gain meaning from it \cite{DBLP:journals/corr/abs-1801-04971}.
 
\section{Open problems} \label{sec:openproblems}
In this survey, we have organized the literature into a framework that reflects the high-level steps necessary to implement a dataset search system. We have considered current research explicitly targeting dataset search challenges. In this section, we discuss several cross-cutting themes that need to be explored in greater detail to advance dataset search. 

Issues of discoverability of open data were recognized by the European Commission which oversees the process of the data publishing within Europe. In $2011$ they defined six barriers that challenge the reuse and true openness of data, which also apply to dataset search \cite{eucommisionmemo:2011}:

\begin{itemize}
\item A lack of information that certain data actually exists and is available
\item A lack of clarity of which public authority holds the data
\item A lack of clarity about the terms of re-use
\item Data which is made available only in formats that are difficult or expensive to use
\item Complicated licensing procedures or prohibitive fees
\item Exclusive re-use agreements with one commercial actor or re-use restricted to a government-owned company.
\end{itemize}

In addition to these challenges, we identify several additional problems that need attention.

\subsection{Query languages: moving beyond keywords}
Existing dataset search systems, whether it is Google's Dataset Search or vertical engines such as those used within data repositories, reuse query languages and concepts from information retrieval. Information needs are expressed via keyword queries, or, in the case of faceted search, via a series of filters modelled after metadata attributes such as domain, format or publisher. Studies in tabular search point to the need for alternative interfaces, which allow users to start their search journey with a table and then add to it as they explore the results. In addition to having different ways to capture information needs, it would also be beneficial to provide query languages that are able to combine information adaptively across multiple tables. This would be especially useful for tasks such as specifying data frames or generating comprehensive data-driven reports \cite{DBLP:journals/corr/abs-1811-06303}. 

This connects dataset search to the area of text databases~\cite{jain2007sql} and the deep web. However, much of that work has looked at verticals instead of search across datasets coming from multiple domains. The problem here is to be able to identify relevant tables for the input query, join them appropriately, and do subsequent query processing. 

Existing research has primarily focused on structured queries (SQL, SPARQL) over the metadata of the datasets, without considering the actual content of the dataset. There is thus a need for richer query languages that are able to go beyond the metadata of datasets and are supported by indexing systems. Our understanding of the level of expressiveness of these languages is still fairly limited. The W3C CSV on the Web working group \cite{Tennison:16:CWA} has made a proposal for specifying the semantics of columns and values in tables, but the approach requires mappings, which are typically specified manually. 

\subsubsection{Entity-centric search building blocks}
Entity-centric search naturally fits within the needs of dataset search. Datasets themselves are often built up of entities, and as such need the ability to specify as a query an entity, set of entities, or type of entity. Moreover, the notion of similarity \cite{Zhang2018AdHT} among entities should be expanded so that the entities themselves are not the focus of the match, but the number of similarities within the dataset.

\subsubsection{Database building blocks}
Querying datasets will likely require new adaptations to query languages and methods. In addition to the exploration of a structured query language that can operate over datasets natively, other mechanisms to define queries should be explored. For instance, the overlap of programming languages and database query languages in which programming language concepts are used to define queries over databases with different levels of capabilities \cite{seco:2017} or over MapReduce frameworks \cite{fegaras:2017}, could be one such rich area to explore. 

\subsubsection{Tabular Search building blocks}
Tabular search provides an interesting view on the potential query language requirements for dataset search, where instead of keywords, the input is a table itself. This also makes novel user interfaces possible, for example, to provide assistance during the creation of spreadsheets \cite{Zhang:2018:SES:3209978.3210219}.

\subsection{Query handling: Differentiated access}
Most dataset search systems today either work within the confines of a single organization or on publicly available datasets that publish metadata according to a specified schema. However, there is demand to be able to pool information stemming from different organizations, for example, to be able to build cohorts for health studies from across clinical studies \cite{Cui2018,Mork2010}. Providing such {\em differentiated access} is critical for the emerging notion of {\em data trusts},\footnote{\url{https://theodi.org/article/what-is-a-data-trust/}} which provide the legal, technical and operational structures to share data between organizations. 

We must facilitate an organizational as well as technical space to share data between both public and private entities. Thus, there are critical issues to be solved with respect querying over datasets with differing legal, privacy and even pricing properties. Without being able to search over these hidden datasets, access to a majority of data will be prevented. Here, aspects of using the provenance of data could be leveraged at query time \cite{wylot2017storing}. We note that this is not just an issue for private data. Public data also has different properties (e.g. licenses) that users want to effectively integrate in their searches. 

At an implementation level, further investigation into integrating security techniques in the query handling process is necessary. For example, searching over encrypted datasets \cite{kumar:2018,bakshi:2018} or using digests to minimize disclosure while still enabling search \cite{Mork2010}. All of this must be done while also considering that the demands of reuse may change the underlying requirements and bottlenecks of query processing \cite{Galakatos:2017:RRA:3115404.3115418}.

\subsubsection{Information Retrieval building blocks}
In the context of dataset retrieval the basic concepts supporting  general web search are not sufficient, which indicates a need for more targeted approach for dataset retrieval, treating it as a unique vertical \cite{Cafarella:2008:WEP:1453856.1453916,Gonzalez:2010:GFT:1807167.1807286}. 

\subsubsection{Database building blocks}
The relational algebra that underpins our processing within a database \cite{codd1972relational}, has no equivalent yet in dataset search. Recently, Apache released information about the query processing system used for many of the Apache products including Hive and Storm, and  \cite{Begoli:2018:ACF:3183713.3190662} investigated how the relational algebra can be applied to data contained within the various data processing frameworks in the Apache suite. Alternatively, other recent work in query processing attempts to handle\newline non-relational operators via adaptive query processing \cite{kaftan:2018}.

Techniques such as those found in \cite{Peng:2018:ACA:3183713.3183747} suggest using a hybrid version of approximate query processing over samples and precomputation. Solutions such as ORCHESTRA \cite{Ives:2008:OCD:1462571.1462577} that were built to manage shared, structured data with changing schemas, cleaning, and queries that utilize provenance and annotation information (discussed in more detail below) need to be adapted to the dataset search problem.  Other work from the probabilistic database area could also be of assistance. For instance \cite{dylla:2013} calculates the top-k results for queries over a probabilistic database by taking into account the lineage of each tuple. This usage of provenance to influence the overall ranking of the end result could inform dataset ranking.

Focusing on constructive dataset search, in which datasets are generated on-the-fly based on a user's needs and query, the work in data integration is particularly important. Querying sources in an integrated fashion 
\cite{halevy2001,konstantinidis2011} becomes a foundational component of constructive dataset search.

\subsection{Data handling: extra knowledge}
In order to support the differentiated access and advanced exploratory interfaces articulated above, dataset search engines will need to become more advanced in their ingestion, indexing and cataloging procedures. This problem divides into two areas:  incorporation of external knowledge in the data handling process and better management and usage of dataset-intrinsic information.

\textbf{Incorporating external knowledge}, whether \\ through the use of domain ontologies, external quality indicators or even unstructured information (i.e. papers) that describe the datasets, is a critical problem. A concrete example of this problem:  many datasets are described through code books that are written in natural language. These datasets are nearly useless without integration of external information about the codebooks themselves.  

\textbf{Utilizing dataset-intrinsic information}, is necessary to more fully capture the richness of each dataset, and allow users to express a richer set of criteria during search. 
Within this space, there are open problems related to data {\em pre-processing}. How to do quality assesment on the fly? What kinds of indexes around quality need to be created? Moving beyond quality, in general, the automatic creation and maintenance of metadata that describes datasets is difficult. Users rely up on metadata to chose appropriate datasets. Open problems for metadata include:
\begin{enumerate}
    \item identifying the metadata that is of highest value to users w.r.t. datasets;
    \item tools to automatically create and maintain that metadata;
    \item automatic annotation of dataset with metadata - linking them automatically to global ontologies.
\end{enumerate}

In addition to pre-processing, current dataset search systems primarily rely on information retrieval architectures (e.g. indexing into ElasticSearch) to index and perform queries. Here, lessons learned from database architectures should be applied. This is particularly the case as we have seen the importance of lessons learned from relational query engines being applied in the case of distributed data environments \cite{armbrust2015spark}. Thus, we think an important open problem is what the most effective {\em architectures} are for dataset search systems. 

\subsubsection{Entity-centric search building blocks}
One can apply the Linked Data paradigm to solve dataset search by converting datasets to RDF and following the full cycle, as described in~\cite{kunze_dataset_2013}. However, for data publishers, it is often still very expensive to execute the full cycle. Furthermore, there is debate on whether certain datasets should have an RDF representation at all, as their original formats are perhaps more suited to the tools that are required for them (e.g. geospatial datasets). A middle-ground solution is to consider datasets as resources and encode only their description in RDF, for example, using the Data Catalog Vocabulary (a W3C recommendation)~\cite{dcat}. Then, the Linked Data cycle can then be applied to these descriptions, ultimately enabling the querying of datasets. The main challenge is the generation and maintenance of these descriptions, with some works tackling the problem of extracting specific properties from specific formats, like ~\cite{neumaier2018enabling} for extracting spatio-temporal properties, and \cite{kacprzak_making_2018} for identifying the numerical properties in CSV tables.

\subsubsection{Database building blocks}
As noted in \cite{bailis2017}, users do not have the `attention' to introspect deeply into large and changing datasets. Instead, we can draw upon several areas of research from the database community, including data profiling and data quality. 

Naumann's recent survey \cite{Naumann:2014:DPR:2590989.2590995} provides a good overview of data profiling activities based on how data-users approach the task, and what resources are available for it.  Of particular note for dataset search is the work on outlier detection \cite{dong2017authenticated,DBLP:conf/sigmod/LiuJ09} as a way to provide indications to an end-user about the scope, spread and variety of a dataset during search. In particular, we note the techniques found in  \cite{Zhang:2016:SEE:3021924.3021928} are interesting for dataset search in that they split a large dataset into many smaller datasets and create an approximate representation of it for more accurate sampling of these sub-pieces.  Finally, \cite{Gao:2018:NDL:3183713.3183746} establishes a tool that can comb through semi-structured log datasets to pull information into multi-layered structured datasets. All of these techniques may aid users in exploring and making sense of dataset. Given that a dataset is by definition a collection of pieces, imputation of missing pieces needs greater scrutiny. As discussed in Section \ref{sec:researchsurvey}, imputation efforts are underway \cite{ahmadov2015towards,BISCHOF201822,li2014web,tang2017efficient} but draw heavily from web techniques. The imputation methods from the data management community should be considered.

The work on profiling  contains expressions of data cleanliness and coverage, completeness and consistency. These properties are classic data quality metrics, and help the user form a picture of whether the data is fit for use. Automatic understanding of data quality in order to either populate metadata or answer metadata queries in a lazy manner will require techniques that can automatically determine complex datatypes such as \cite{Yan:2018:STL:3183713.3196888}. Currently, though, the research in each of these areas has been focused on its relationship to describing or working within a specific artifact, not as a component for a search. To do this, the structures and content for each area need to be computable in a timely manner and presented in a way that can be taken advantage of by a search system. For instance, data quality is a traditionally resource expensive task that is often domain-specific. Generic, albeit possibly less accurate methods must be developed to compute data quality estimations that can be accessed and used during search \cite{Chapman:2016:CLD:2888577.2834123,missier:2006}. 

In order to facilitate understanding of the contents of a dataset, summarization can be used, as done in \cite{Orr:2017:PDS:3115404.3115419} over probabilistic databases.  Provenance, another tools that could help users understand a dataset, has an unsolved problem of moving across granularity levels. A tuple within a dataset may have provenance associated with it, as may the table, and the entire dataset itself. The challenge is in understanding how the aggregation of tuple-provenance would affect the search results compared to dataset-provenance. Finally, using annotations to improve the data \cite{Ibrahim:2015:PAM:2723372.2749435} will be needed. Interesting extensions could include using user feedback to facilitate ranking of datasets based on the searcher's criteria, or utilizing the context under which the annotations were created to change how annotations impact ranking. 

\subsubsection{Hidden/Deep Web building blocks}
An inherent challenge in dataset search over the web is to be able to identify particular resources as datasets of interest (and ignore, for example, natural language documents). This challenge will be also present in any forthcoming approach in searching for datasets on the deep web. Moreover, any such approach will build on some combination of the two main directions for surfacing deep web data. Building vertical engines for the hidden web has the difficulties of pre-defining all interesting domains, identifying relevant forms in front of datasets on the web and investigating automatic (or semi-automatic) approaches to create mappings; a task which seems extremely hard on a web scale. Hence, learning/computing web form inputs might be the option of choice. Nevertheless, in cases where there are complex domains that involve many attributes and involved inputs, e.g., airline reservations, when the datasets change frequently, e.g., financial data, or when forms use the http POST method \cite{madhavan2008google} virtual integration remains an attractive direction.

\subsubsection{Tabular Search building blocks}
The majority of work in tabular search addresses web tables, not uploaded datasets. These tables have the benefit of generally being better described and often general-knowledge related, e.g., column names are human readable and not codes, or the tables are embedded in larger documents (e.g. HTML tables). In addition, a majority of work treats what are termed `entity-centric tables', which are tables in which each row represents a single entity. Datasets can be much more general, for example, containing multiple tables in one file.

\subsection{Result presentation: interactivity}
As previously discussed, existing data search systems follow similar approaches to search showing a ranked list of search results with some additional faceted searching in place. At a tactical level, ranking approaches specifically tailored to dataset search should be developed. Importantly, this should take into account the kinds of rich indexes suggested in the prior section. Here, the challenges are that typical approaches to improving ranking from information retrieval such as learning to rank are difficult given that many data search engines do not have the kind of level of user traffic needed for learning to rank algorithms \cite{van-gysel-neural-2018}.  In addition, the integration of dataset search and entity search is an important open problem. For example, when searching for a chemical could you also display associated data and what that data should be. 

Beyond standard search paradigms, supporting conversational search over data and embedding search into the actual data usage process deserves significant attention, particularly since dataset search is often needed in the context of a variety of tasks \cite{Stonebraker2018}.

\subsubsection{Information Retrieval building blocks}
As pointed out by Cafarella et al. \cite{Cafarella:2008:WEP:1453856.1453916} structured data on the web is similar to the scenario of ranking of millions of individual databases. Tables available online contain a mixture of structural and related content elements which cannot easily be mapped to unstructured text scenarios applied in general web search. Tables lack the incoming hyperlink anchor text and are two-dimensional - they cannot be efficiently queried using the standard inverted index. For those reasons PageRank-based algorithms known from general web search are not applicable to the same extend to the dataset/table search, particularly as tables of widely-varying quality can be found on a single web page.

Search for datasets is often complex and shows characteristics of exploratory search tasks, involving multiple queries, iterations and refinement of the original information need, as well as complex cognitive processing \cite{DBLP:conf/chi/KoestenKTS17}. There are many possible reasons that users have diverse interaction styles, from context and domain specificity  \cite{GregoryGCSW17} to uncertainty in the search workflow itself \cite{Boukhelifa:2017:DWC:3025453.3025738}. It is important to note that users  have different interaction styles with respect to 'getting the data'. These interactions range from question answering to "data return" to exploration \cite{GregoryGCSW17,DBLP:conf/chi/KoestenKTS17}. From an interaction perspective, dataset search is not as advanced as web or document search. Contextual or personalized results, which are common on the web \cite{white2009predicting} are practically non-existent for dataset search. Additionally, dataset search relies on limited metadata instead of looking at the dataset itself. While many classifications for information seeking tasks exist \cite{blandford2010interacting}, there is no widely used classification of dataset  information seeking tasks yet. 

\subsubsection{Database building blocks}
 Provenance \cite{Buneman:2006:PMC:1142473.1142534,Green:2007:PS:1265530.1265535,DBLP:journals/vldb/HerschelDL17,wylot2017storing} is likely to be a key element in assisting the user in choosing a dataset of interest. Until now, provenance has been used to facilitate trust in an artifact \cite{CURCIN20171,dai:2008} or automatically estimate quality \cite{huynh:2018}. New methods must be developed to facilitate translation of this large graph into a format that a user who is evaluating whether or not to use a dataset can interpret and utilize \cite{chapman:2011}. The logic and possible new operators behind dataset search will open up new areas for determining why and why  not to consider provenance of the dataset query results themselves \cite{Chapman:2009:WHY:1559845.1559901,DBLP:journals/vldb/HerschelDL17,lee:2017}.

The presentation of data models has been a topic in database literature \cite{DBLP:conf/sigmod/JagadishCEJLNY07} as well as exploration strategies of result spaces beyond the 10 blue links paradigm. For instance, the use of sideways and downwards exploration of web table queries by \cite{Chirigati:2016:KEU:3021924.3021935}. 
 Challenges and directions for search results presentation and data exploration as part of the search process are discussed on a mostly speculative basis in literature, and include representing different types of results in a manner that express the structure of the underlying dataset (tables, networks, spatial presentations,etc) \cite{DBLP:conf/sigmod/JagadishCEJLNY07}.  

An overview of search results can enhance orientation and understanding of the information provided \cite{rieh2016towards}, which allows to get an awareness of the dataset result space as a whole. Making a large set of possible results more informative to the user has been explored for databases \cite{Wen:2018:ISE:3275366.3284965}. At the same time being able to investigate the dataset on a column, row and cell level to match both process and content oriented requirements on the search result can be necessary \cite{Pimplikar:2012:ATQ:2336664.2336665,Tennison:16:CWA}. 

Within the scope of constructive dataset search,  the work of \cite{Wu:2018:DCG:3183713.3196910} is essential to appropriately annotate and cite the results of queries.

In the next section, we discuss one foundation that is crucial for addressing these open problems, benchmarks.

\section{The Road Forward: Benchmarks} \label{sec:route}

% "The rise of machine learning challenges, with large, common test sets ensures that results are more directly comparable,"  \cite{witten:2011}. 
% Benchmarks exist in all areas related to dataset search, from RDF \cite{Pan2018}, to IR \cite{}, etc. Even machine learning communities are establishing benchmarks, e.g. by using Kaggle datasets, and the features that were actually used by the winners of Kaggle Challenges \cite{PLiu:2018:MBM:3231751.3242940}. This does not yet exist for Dataset search yet, and should be a priority to facilitate research in this area. 

One of the most widely recognized problems of dataset search is the lack of benchmarks. For instance, the BioCADDIE project, which attempts to index for discovery scientific datasets, has a pilot project to recommend appropriate datasets to users based on similar topic, size, usage, user background and context \cite{biocaddePilotProject}. In order to do this, the pilot participants are creating a topic model across scientific articles, and using user query patterns to identify similar users. While this is an interesting start, and acknowledges that there are a myriad of overlapping concerns that impact dataset search, from content through user's ability, there is no way yet to measure whether the solution works. For this, a clear benchmark is needed. In this section we will outline the state of the art with respect to the evaluation of different parts of the dataset search pipeline, which were discussed earlier in this work. 

Step one is identifying the set of metrics that are appropriate to dataset search. Do they mimic the online and offline metrics of information retrieval? At first blush, session abandonment rate, session success rate and  zero result rate from information retrieval online metrics appear relevant, while click-through rate may need some adjustment for the context of datasets.  Meanwhile, most of the offline metrics, from the set of precision-based metrics, to recall, fall-out, discounted cumulative gain, etc. are obviously still necessary. 

However, there are  dataset-specific metrics that may need to be considered. For instance, ``completeness" could be an interesting new metric to consider. Many tasks involving datasets require the stitching of several datasets to create a whole that is fit for purpose. Is the right set, that creates a ``complete" offering returned? How do we measure that the appropriate set of datasets for a given purpose were returned. For instance, in the context of information retrieval on an Open Data Platform, \cite{kacprzak2018} found that some user queries require multiple datasets which are equally relevant in opposition to a ranked result list of resources with single resource per rank. The question of how such result list should be returned to the user remains open, and creates an interesting case within benchmark creation.

The availability of benchmarks upon which solution across the query processing pipeline for dataset search can be tested is essential. Any benchmark created for dataset search needs to, explicitly or implicitly, highlight the relationships that exist between the user, the task at hand and the properties of the dataset or it’s metadata. %Unlike classic web retrieval, there is an added dimension for dataset search. It is no longer enough for a user to find the information appropriate; for dataset search, the user and the task must be satisfied and the result list presented to the user must be understandable.
Unlike classic web retrieval, there are added dimensions for dataset search. It is no longer enough for a user to find the information appropriate; for dataset search, the user and the specific task requirements must be satisfied. The result list presented to the user must be understandable and explorable, due to the added complexity of interpreting and using data.

Several benchmarks have already been created that cover tasks related to dataset search. These benchmarks include: managing RDF datasets \cite{Pan2018}; information retrieval over Wikipedia tables \cite{Zhang2018AdHT}; assignment of semantic labels to web tables \cite{DBLP:conf/wims/RitzeLB15}. Further efforts in this area needed in order to truly understand and make progress on the underlying technology.

% There are burgeoning steps in this area. For instance, Zhang et al. provide an open corpus of tables designed for measuring performance \TODO{on this task - what task?} \cite{Zhang2018AdHT}. 

\section{Conclusions}
The topic of data-driven research will only grow; we are at the start of a journey in which datasets are used for analysis, decision making and resource optimization. Our current needs for Dataset Search require us to give due attention to this problem. The current state-of-the-art is focused on tuple, document or webpage. Datasets are an interesting entity to themselves with some properties shared with documents, tuples and webpages, and some unique to datasets. 

In this work, we highlight that dataset search can be achieved through two different mechanisms: 1. issue query, return dataset; 2. issue query, build dataset. However, dataset search itself is in its infancy. Techniques from many other fields, including databases, information retrieval, and semantic web search can be applied towards the problem of dataset search. The creation of an initial service, Google Dataset Search, that allows for automatic indexing of datasets, and Google-style search over that indexed information marks this problem as important. Moreover, it highlights the research that still needs to be performed within the dataset retrieval domain, including: formal query language(s), dealing with social and organizational restrictions when processing a query, providing additional information to support query processing, facilitating user exploration and interaction with a result set made up of datasets. This is an exciting time with respect to dataset search, in which there is a high need for datasets of all sorts, combined with burgeoning tools for dataset search, like Google Dataset Search, that provide the necessary infrastructure. However, further research is needed to fully understand and support dataset search.

% originally written paper
%\input{introduction.tex}
%\input{currenttech.tex}

%\input{challenges.tex}
%\input{conclusions.tex}

%\begin{acknowledgements}
%\end{acknowledgements}

\bibliographystyle{spmpsci}      % mathematics and physical sciences
\bibliography{datasetret}   % name your BibTeX data base

\end{document}